\documentclass[english,eqsecnum,prd,aps,nofootinbib,superscriptaddress,longbibliography,tightenlines,12pt]{revtex4-2}

\usepackage[T1]{fontenc}
\usepackage[latin9]{inputenc}
\usepackage[dvipsnames]{xcolor}
\usepackage{babel}
\usepackage{mathtools}
\usepackage{amsmath}
\usepackage{graphicx}
\usepackage{float}
\usepackage{amssymb,accents}
\usepackage{mathrsfs}
\usepackage{cancel}
\usepackage{comment}
\usepackage{physics}
\usepackage[unicode=true,pdfusetitle,
 bookmarks=true,bookmarksnumbered=false,bookmarksopen=false,
 breaklinks=true,allcolors=blue,backref=false,colorlinks=true]
 {hyperref}
 
 \usepackage{xcolor}

 \usepackage[normalem]{ulem} 

\begin{document}

\title{Post-Newtonian Effects in Compact Binaries with a Dark Matter Spike: A Lagrangian Approach}

\author{Diego Montalvo}
\email{montalvo@cita.utoronto.ca}
\affiliation{Canadian Institute for Theoretical Astrophysics, University of Toronto, 60 St. George Street, Toronto, ON M5S 3H8, Canada}
\affiliation{Department of Physics, University of Toronto, 60 St. George Street, Toronto, ON M5S 1A7, Canada}

\author{Adam Smith-Orlik}
\email{asorlik@yorku.ca}
\affiliation{Department of Physics and Astronomy, York University~\\
 4700 Keele Street, Toronto, Ontario M3J 1P3 Canada}

\author{Saeed Rastgoo}
\email{srastgoo@ualberta.ca}
\affiliation{Department of Physics, University of Alberta, Edmonton, Alberta T6G 2G1, Canada}
\affiliation{Department of Mathematical and Statistical Sciences, University of Alberta, Edmonton, Alberta T6G 2G1, Canada}
\affiliation{Theoretical Physics Institute, University of Alberta, Edmonton, Alberta T6G 2G1, Canada}

\author{Laura Sagunski}
\email{sagunski@itp.uni-frankfurt.de}
\affiliation{Institute for Theoretical Physics, Goethe University, 60438 Frankfurt am Main, Germany}

\author{Niklas Becker}
\email{nbecker@itp.uni-frankfurt.de} 
\affiliation{Institute for Theoretical Physics, Goethe University, 60438 Frankfurt am Main, Germany}

\author{Hazkeel Khan}
\email{hazkeelk@my.yorku.ca}
\affiliation{Department of Physics and Astronomy, York University~\\
 4700 Keele Street, Toronto, Ontario M3J 1P3 Canada}

\date{\today}

\begin{abstract}
We apply the Lagrangian method to study the post-Newtonian evolution of a compact binary system with environmental effects, including a dark matter spike, and obtain the resulting gravitational wave emission. This formalism allows one to incorporate post-Newtonian effects up to any desired known order, as well as any other environmental effect around the binary, as long as their dissipation power or force formulae are known. In particular, in this work, we employ this method to study a black hole--black hole binary system of mass ratio $10^5$ by including post-Newtonian effects of order 1PN and 2.5PN, as well as the effect of relativistic dynamical friction. We obtain the modified orbits and the corresponding modified gravitational waveform. Finally, we contrast these modifications against the LISA sensitivity curve in frequency space and show that this observatory can detect the associated signals.
\end{abstract}

\maketitle

\section{Introduction\label{sec:Intro}}

The first direct detection of gravitational waves (GWs) by the LIGO/Virgo collaboration has opened up a new window into the universe~\cite{TheLIGOScientific:2016src}. Mergers of compact binary 
objects, such as black holes and neutron stars, provide unprecedented precision tests of general relativity and matter at its highest densities. There are also plans for space-based GW observatories such as The Laser Interferometer Space Observatory (LISA)~\cite{LISA:2017pwj}, Taiji~\cite{10.1093/nsr/nwx116}, and TianQuin~\cite{TianQin:2015yph}. These will be able to observe GWs at lower frequencies and, thus, to observe mergers of massive binary black holes and intermediate and extreme mass ratio inspirals (EMRIs/IMRIs). These systems are considered to be a rich source of signatures associated with various new and fundamental physics phenomena~\cite{LISA:2022kgy, LISACosmologyWorkingGroup:2022jok, Addazi:2021xuf,AlvesBatista:2023wqm, Garcia-Chung:2023oul, Garcia-Chung:2022pdy, Garcia-Chung:2020zyq} and, in particular, dark matter (DM).
The standard cosmological $\Lambda$CDM model predicts the existence of such dark matter: a cold, collisionless massive particle that has so far eluded our direct detection efforts~\cite{Bertone:2004pz}. Utilizing this new window into the universe, GW probes of dark matter have been gaining traction in recent years~\cite{Bertone:2018krk}.

When intermediate mass black holes (IMBHs) undergo adiabatic growth within DM halos, overdensities of DM, so-called DM \textit{spikes}
, can form~\cite{Gondolo:1999ef, Sadeghian:2013laa}. We can find extensive work performed on modeling the spacetime of these structures, as well as their analytical forms given certain physical conditions, in Refs.~\cite{maedaEinsteinClusterCentral2024, shenAnalyticalModelsSupermassive2024, shenClassAnalyticalModels2024, Cardoso:2021wlq}.  The presence of these DM spikes can affect an inspiraling object as part of an IMRI. This was first explored in~\cite{Eda:2013gg, Eda:2014kra}, where the DM spike interacts with the compact object through dynamical friction. This results in a faster inspiral compared to an inspiral in a vacuum and would be observable as a \textit{dephasing} of the GW signal, possibly detectable by LISA~\cite{Barrau:2014maa, Coogan:2021uqv}.

Additional effects of these DM spikes have been explored in consecutive works, such as the effects of accretion of the DM spike~\mbox{\cite{Macedo:2013qea, Yue:2017iwc}}, eccentric orbits inside these DM spikes~\mbox{\cite{Yue:2019ozq, Cardoso:2020iji, Becker:2021ivq, Becker:2022wlo}}, periastron precession~\cite{Dai:2021olt}, the halo feedback mechanism~\cite{Kavanagh:2020cfn, Coogan:2021uqv}, relativistic corrections to dynamical friction and spike distribution~\cite{maedaEinsteinClusterCentral2024, Speeney:2022ryg}, and DM spikes around primordial black holes~\cite{Cole:2022ucw}. These works all explore different effects that need to be combined in the end, as precise waveforms are needed to find them in the LISA data~\cite{Arnaud:2006gm}.

In this work, we present a general Lagrangian framework that can easily incorporate post-Newtonian (PN) corrections, dark matter dynamical friction, accretion, and any other orbital or environmental effect in a compact binary system with environment, as long as the mathematical formula of the aforementioned effects and corrections are known in the form of a dissipative power or a force. These effects are then formulated as generalized forces, and the Euler--Lagrange equations of motion yield the modified orbits, from which the waveform of the gravitational waves emitted by the binary system can be derived. 
Thus, this framework is different from other approaches that assume the quasi-adiabatic inspiral in their computations. 

The structure of this paper is as follows. In Section~\ref{Sec:DmHalo}, we present the system we are considering applying our Lagrangian formulation to and describe the model used for the DM spike profile. In Section~\ref{Sec:Lagrangian}, we present the main proposal of the paper, namely, the introduction of a Lagrangian formulation that can incorporate all the PN corrections up to any order, as well as other effects such as dynamical friction due to DM and mass accretion by the binary system.
Section~\ref{Sec:Orbits-GWs} is dedicated to studying the system detailed in Section~\ref{Sec:DmHalo}, using the Lagrangian formalism of  Section~\ref{Sec:Orbits-GWs}. There, we compute the orbits from which we derive the waveform of the GWs emitted by the compact binary system surrounded by a spike. We also compare the results with no-DM and no-PN cases and, by an analysis in the frequency space, show that these modifications to the GWs can be observed by LISA. Finally, in Section~\ref{sec:conclusion}, we summarize our work and make some concluding remarks about the potential further applications of the Lagrangian framework we present in this paper.


\section{Dark Matter Halo\label{Sec:DmHalo}}
\subsection{Spike Profile}

We consider a Schwarzschild black hole (BH) with mass $m_1$ that grows adiabatically and forms a surrounding DM spike $\rho_{\rm DM}(r)$ that concentrates DM from an initial Navarro--Frenk--White (NFW) profile~\cite{Navarro_1997}. A second smaller compact object of mass $m_2 \ll m_1$ is on an approximately Keplerian orbit around the BH and experiences dissipative forces from gravitational wave emission and dynamical friction from the DM halo, which cause an inspiral due to the loss of orbital energy. 

During the adiabatic growth of the central BH, a DM halo can contract and form a spike, resulting in large DM densities close to the BH horizon. The DM density profile describing such a spike was derived first in a semi-relativistic Newtonian manner~\cite{Gondolo:1999ef} and later in a fully relativistic manner 
~\cite{Sadeghian:2013laa}. The fully relativistic model predicts that the DM density vanishes at $2R_s$ compared to $4R_s$ predicted by the semi-relativistic treatment, where $R_s$ is the Schwarzschild radius of the BH. Furthermore, the central densities of the relativistic DM spike can be significantly higher compared with the semi-relativistic case. Higher DM densities can have a significant impact on the rate of inspiral and, hence, the gravitational wave signal; therefore, we elect to use the fully relativistic model for our DM spike.

To model a relativistic DM density spike we follow the effective scaling function, Equation~(7) in~\cite{Speeney:2022ryg}, given as
\begin{equation}
    \label{eq:rho_relativistic}
    \rho_{\rm DM}(r) = \bar{\rho} 10^{\delta}\left(\frac{\rho_{0}}{\rm 0.3~GeV/cm^3}\right)^{\alpha}\left(\frac{m_1}{10^6 \rm M_{\odot}}\right)^{\beta}\left(\frac{a}{20~ \rm kpc}\right)^{\gamma}, \; 
\end{equation}
\noindent
with 
\begin{equation}
    \label{eq:rho_prefactor}
    \bar{\rho} = A\left( 1-\frac{4}{\tilde{x}}\right)^{w}\left(\frac{4.17\times 10^{11}}{\tilde{x}}\right)^{q},
\end{equation}
\noindent
where $\alpha$, $\beta$, $\gamma$, $\delta$ are the relativistic NFW parameters generated by comparison with numerically generated curves; $A$, $w$, and $q$ are fit parameters found by fitting to a reference curve with the scale parameters $(\rho_{0}, M_{BH}, a)=(\rm 0.3GeV/cm^3, 10^6 \rm M_{\odot},20~\rm kpc)$; and $\tilde{x}=r/m_1$ (see Table~1 in Ref.~\cite{Speeney:2022ryg})\footnote{Note that $\rho_{\rm DM}(r)$ is a valid fit when $r\ll a$ and $0.01~\rm kpc \leq a$. }.

For the inspiral process, the DM halo is assumed to be static. This assumption ignores the effects of dynamical friction on the DM halo itself, also referred to as halo feedback, which can be significant for binary systems with mass ratios less than $10^{5}$, as shown in Ref.~\cite{Kavanagh:2020cfn}. Nevertheless, in this work, we first would like to check the consistency of applying the Lagrangian formulation to explore the effects of the post-Newtonian correction terms in the environment of a dark matter spike.

In the presence of a DM spike, the additional dissipative forces will speed up the inspiral, which is potentially observable in the GW signal. Therefore, it should, in principle, be possible to map the spike density. The application of match-filtering analyses such as the one in~\cite{Cole:2022ucw} are powerful tools for GW detections and parameter inference. It is, however, necessary to utilize suitable GW templates that, by taking into account a broader class of physical effects such as a relativistic DM spike or orbital evolution beyond the Newtonian regime, could improve the possibilities of potential detection. We consider the evolution of the binary system slightly before GW emission enters the lower end of the detector's band. For our binary mass considerations, this coincides with regions of high density from the DM spike at a distance from the central BH of about a hundred times the innermost stable circular orbit, $r_{\rm ISCO}$, defined as  
\begin{equation}
    \label{eq: risco}
    r_{\rm ISCO} = 3R_{\rm s} \;.
\end{equation}

\subsection{System Parameters}
\begin{figure}[H] 
  \includegraphics[width=0.85\textwidth]{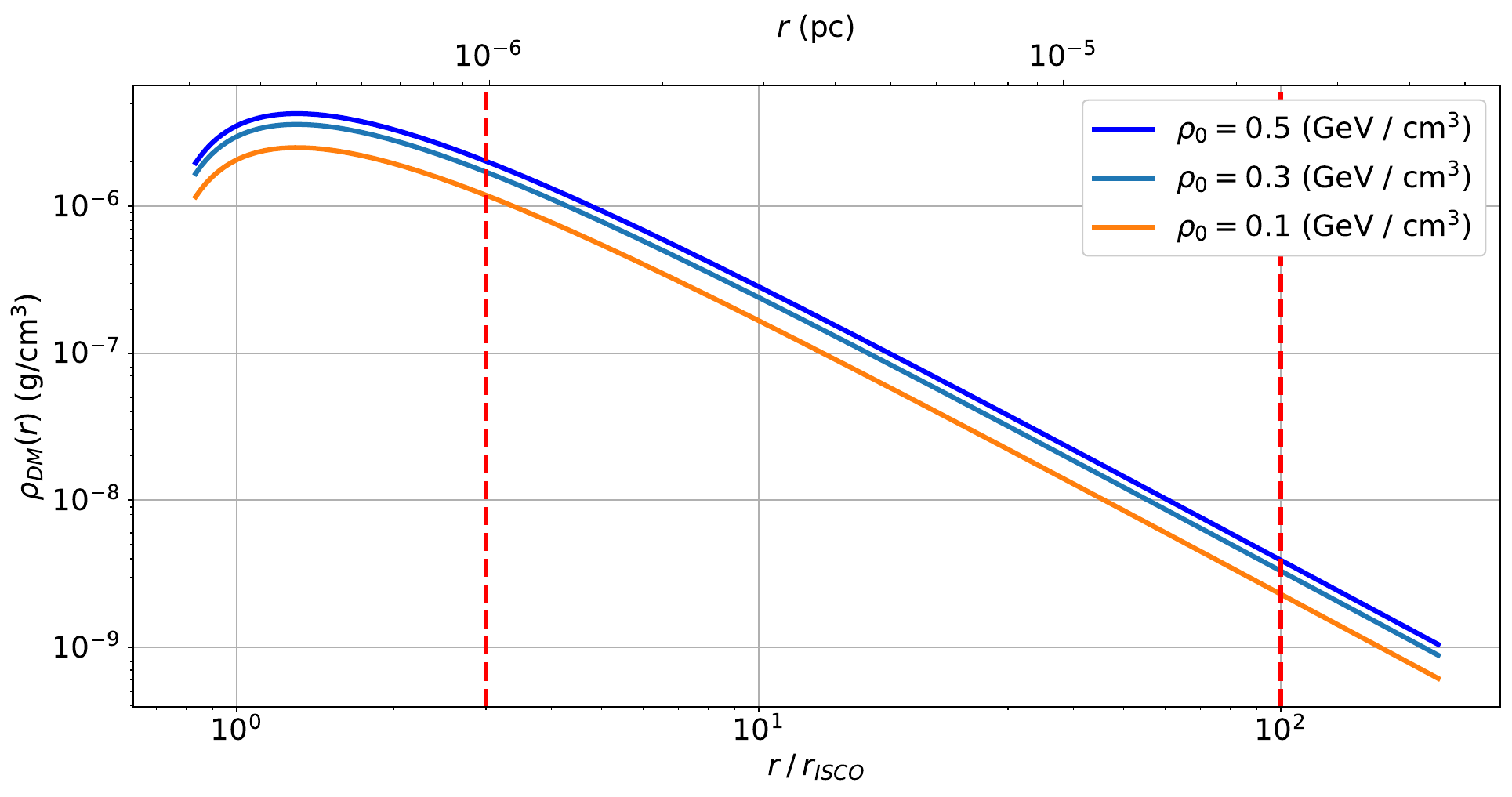} 
    \caption{{Figure of the DM profile used under the relativistic corrections in~\cite{Speeney:2022ryg} for \mbox{$\rho_0 = (0.1, 0.3, 0.5)\;\rm{GeV/cm^{3}}$}.} The dashed red lines represent the stages of the evolution. The right line is situated at the start of the evolution at $100r_{ISCO}$, and the left line is where we terminate the inspiral at $3r_{SICO}$. These ranges correspond to observable GW frequency bands for LISA, as well as the range of validity of PN corrections and final stages of the inspiral.}
  \label{fig:density}
\end{figure}

In this work, we employ the static halo approximation, for which, following in the spirit of~\cite{Kavanagh:2020cfn, Speeney:2022ryg}, we focus our study on a central mass of $10^6 M_{\odot}$\footnote{Which matches with the scale $M_{\rm BH}$ in Equation~\eqref{eq:rho_relativistic}} with a $10 M_{\odot}$ companion. This produces a mass ratio of $10^5$, which is high enough such that a static halo model can be used as halo feedback, which 
becomes important at lower mass ratios, as well as ranging to the higher IMRI regime where most EMRIs are expected to have shed their dark matter halos~\cite{Kavanagh:2020cfn}. To emphasize the sensitivity of the inspiral due to the DM density, we present three DM spikes in Section~\ref{Sec:Orbits-GWs} with varying scale densities, $\rho_0$, i.e.,\linebreak   $\rho_0 = (0.1, 0.3, 0.5)\;\rm{GeV/cm^{3}}$\footnote{Corresponding to the three DM scale densities tested in Ref.~\cite{Speeney:2022ryg}.}. For simplicity, we match the scale radius to the reference value, i.e., $a=20\; \rm{kpc}$. Moreover, we assume a distance of $1\; \rm{Mpc}$  from the Earth to the binary and a total inspiral distance for the companion to be from $100 \: r_{\rm ISCO} $ to $3 \: r_{\rm ISCO}$, which is sufficient to show the effects of PN corrections, DM friction, while also being inside the range of LISA's sensitivity band. Note that for these mass ranges (i.e., central black hole standing at $\sim$$10^6 M_{\odot}$), one may start to consider a self-force approach to model the waveform. However, we are considering the case where the evolution takes place at a relatively large distance of $\sim$$100r_{\text{ISCO}} - 3r_{\text{ISCO}}$. With these conditions, the companion is situated far enough away from the strong field regimes and never exceeds $v/c \sim 0.3$, where PN correction terms are known to converge slowly~\cite{maggioreGravitationalWavesVol2018}. An illustration of the dark matter density profile and the stages of the evolution for our system can be found on Figure~\ref{fig:density}.

\section{Lagrangian Formulation and Equations of Motion\label{Sec:Lagrangian}}
\label{sec:eom}

To incorporate post-Newtonian (PN) corrections and DM effects in our model to be able to compute the orbits with these corrections taken into account, we use the Lagrangian formulation and encapsulate 
PN and DM effects as generalized forces in this method. To see this more clearly, consider a system that is under both $b$ Newtonian forces 
 $\tilde{\mathbf{F}}^{(b)}$
and $l$ non-Newtonian forces $\mathbf{F}^{(l)}$. The Lagrangian of the system can still be written as
\begin{equation}
L=T-V
\end{equation}
where $T$ is the kinetic energy of the system, and $V$ is the potential
energy associated to the Newtonian forces $\tilde{\mathbf{F}}^{(b)}$. However, using the d'Alembert principle, the Euler--Lagrange (E-L) equations of motion can be written as\footnote{We use Einstein's summation notation and the indices are raised
and lowered with the flat Euclidean metric $\delta_{ij}=\mathrm{dig}(1,1,1)$.}
\begin{equation}
\frac{d}{dt}\left(\frac{\partial L}{\partial\dot{q}^{i}}\right)-\frac{\partial L}{\partial q^{i}}=\sum_{l}Q_{i}^{(l)}\label{eq:EL-1}
\end{equation}
where $Q_{i}^{(l)}$ represents generalized forces corresponding to non-Newtonian forces $\mathbf{F}^{(l)}$, i.e.,
\begin{equation}
Q_{j}^{(l)}=\mathbf{F}^{(l)}\cdot\frac{\partial\mathbf{r}}{\partial q^{j}}=F_{i}^{(l)}\frac{\partial x^{i}}{\partial q^{j}}.\label{eq:Qnc}
\end{equation}
Here,  $\mathbf{r}$ is the Cartesian position vector of the object (or subsystem), and $q^{j}$ is its generalized coordinates, where in 3D, $j=1,2,3$. We can adopt $q^{j}$ to spherical coordinates such that $q^{1}=r,\,q^{2}=\theta,\,q^{3}=\phi$, and thus,
\begin{equation}
\mathbf{r}=\left(x^{1},x^{2},x^{3}\right)=\left(x,y,z\right)=\left(r\cos\left(\phi\right)\sin\left(\theta\right),r\sin\left(\phi\right)\sin\left(\theta\right),r\cos\left(\theta\right)\right).
\end{equation}
If the motion is restricted to a 2D plane, for example, in the case of a Keplerian orbit of a planet with $\theta=\frac{\pi}{2}$, then
we can effectively work with just two coordinates  $q^{1}=r,\,q^{2}=\phi$, and thus,
\begin{equation}
\mathbf{r}=\left(x^{1},x^{2}\right)=\left(x,y\right)=\left(r\cos\left(\phi\right),r\sin\left(\phi\right)\right).\label{r_in_terms_of_q_polar}
\end{equation}
For such a 2D system with $l$ non-Newtonian forces, Equation (\ref{eq:EL-1})
becomes
\begin{align}
\frac{d}{dt}\left(\frac{\partial L}{\partial\dot{r}}\right)-\frac{\partial L}{\partial r}= & \sum_{l}Q_{r}^{(l)},\label{eq:EL-Gen-1}\\
\frac{d}{dt}\left(\frac{\partial L}{\partial\dot{\phi}}\right)-\frac{\partial L}{\partial\phi}= & \sum_{l}Q_{\phi}^{(l)}.\label{eq:EL-Gen-2}
\end{align}
As mentioned above, in our model, we deal with both the PN corrections and the so-called dynamical
friction effects. The latter stem from the interaction of the orbiting object (with lower mass) with its local DM environment and have nothing to do with friction in an electromagnetic sense.

In what follows,  we add the conservative 1PN term and both  dissipative 2.5PN and the dynamical friction terms, which are velocity-dependent, to the E-L equations of motion as generalized forces.
Among the aforementioned PN terms, we first consider the leading terms that give corrections to the orbit and the energy dissipation through gravitational waves. The first orbit correction term is a conservative term related to the orbital precession, coming in at first order or 1PN~\cite{Will:2011nz,blanchetPN}. The next leading term related to gravitational wave energy dissipation is a non-conservative one and is precisely equivalent to the so-called order 2.5PN correction term~\cite{Will:2011nz}.

\subsection{Newtonian Part of the Euler--Lagrange Equations}

As mentioned above, in our two-body system that is surrounded by a DM spike, in the center of mass (CM) frame, we can write the Lagrangian
as 
\begin{equation}
L=  \frac{1}{2}\mu\dot{\mathbf{r}}^{2}-V(r) \label{eq:L}
\end{equation}
where $\mathbf{r}=\mathbf{r}_{1}-\mathbf{r}_{2}$ is the relative
position of the objects in which indices $1,\,2$ refer to the central
and the rotating objects, respectively, and $\dot{\mathbf{r}}^{2}=v = \dot{r}^{2}+r^{2}\dot{\phi}^{2}$. We denote the mass of the central
more massive object by $m_{1}$ and the lighter orbiting object by
$m_{2}$. Furthermore, the total mass of the two-body
system is denoted by $m=m_{1}+m_{2}$ and its reduced mass by $\mu=m_{1}m_{2}/m$. The potential used in the Lagrangian is the Newtonian gravitational potential energy between the two bodies,
\begin{equation}
V(r)=-G\frac{\mu m}{r} \,.\label{eq:V-Newton}
\end{equation}
Both masses can potentially be time-dependent if one
also considers the accretion effect, by which they absorb DM from the DM spike around the central object.

\subsection{Generalized Forces}

Once we have the Lagrangian, we can write down the left-hand side of the E-L Equations \eqref{eq:EL-Gen-1} and \eqref{eq:EL-Gen-2}. To write the right-hand side of these equations, we need to find the generalized forces. To carry that out, we assume that the corresponding
usual forces themselves have two components: One component is (anti)parallel
to the relative velocity of the objects $\mathbf{v}=\mathbf{v}_{1}-\mathbf{v}_{2}$,
and the second component is (anti)parallel to $\mathbf{r}=\mathbf{r}_{1}-\mathbf{r}_{2}$,
along the line connecting the two bodies. Such a force can be written
in the form
\begin{equation}
\mathbf{F}^{(l)}=F_{r}^{(l)}\hat{\mathbf{r}}+F_{v}^{(l)}\hat{\mathbf{v}}.\label{eq:Fl}
\end{equation}
The corresponding generalized force is
\begin{equation}
Q_{j}^{(l)}=  \mathbf{F}^{(l)}\cdot\frac{\partial\mathbf{r}}{\partial q^{j}}= \frac{F_{r}^{(l)}}{r}\left(x\frac{\partial x}{\partial q^{j}}+y\frac{\partial y}{\partial q^{j}}\right)+\frac{F_{v}^{(l)}}{v}\left(\dot{x}\frac{\partial\dot{x}}{\partial\dot{q}^{j}}+\dot{y}\frac{\partial\dot{y}}{\partial\dot{q}^{j}}\right),
\end{equation}
where we have used $\frac{\partial\mathbf{r}}{\partial q^{j}}=\frac{\partial\dot{\mathbf{r}}}{\partial\dot{q}^{j}}$.
Considering \eqref{r_in_terms_of_q_polar} and the fact that the index $j$ in the above equation takes values $j=r,\phi$, the radial and angular parts
of the generalized force $Q_{j}^{(l)}$ above become (with $v=\sqrt{\dot{r}^{2}+r^{2}\dot{\phi}^{2}}$)
\begin{equation}
Q_{r}^{(l)}=  F_{r}^{(l)}+\frac{F_{v}^{(l)}\dot{r}}{\left(\dot{r}^{2}+r^{2}\dot{\phi}^{2}\right)^{\frac{1}{2}}},\label{eq:QAr}
\end{equation}
and
\begin{equation}
Q_{\phi}^{(l)}= \frac{F_{v}^{(l)}r^{2}\dot{\phi}}{\left(\dot{r}^{2}+r^{2}\dot{\phi}^{2}\right)^{\frac{1}{2}}}.\label{eq:QAphi}
\end{equation}
On the other hand, if we do not know the force itself but know the corresponding dissipative
power $P^{(l)}$, we can write, using \eqref{eq:Fl},
\begin{equation}
P^{(l)}=\frac{dE^{(l)}}{dt}=  \mathbf{F}^{(l)}\cdot\mathbf{v}=F_{r}^{(l)}\dot{r}+F_{v}^{(l)}\left(\dot{r}^{2}+r^{2}\dot{\phi}^{2}\right)^{\frac{1}{2}}.
\end{equation}
where we have used the polar coordinates expressions $\mathbf{v}=\dot{r}\hat{\mathbf{r}}+r\dot{\phi}\hat{\boldsymbol{\phi}}$
and $\mathbf{r}=r\hat{\mathbf{r}}$.
In cases where the force is (anti)parallel to $\mathbf{v}$, i.e.,
when $F_{r}^{(l)}=0$, we obtain
\begin{equation}
F_{v}^{(l)}=\frac{P^{(l)}}{\left(\dot{r}^{2}+r^{2}\dot{\phi}^{2}\right)^{\frac{1}{2}}},
\end{equation}
and (\ref{eq:QAr}) and (\ref{eq:QAphi}) reduce to 
\begin{align}
Q_{r}^{(l)}= & \frac{P^{(l)}\dot{r}}{\dot{r}^{2}+r^{2}\dot{\phi}^{2}},\label{eq:Ql-r-gen-no-r}\\
Q_{\phi}^{(l)}= & \frac{P^{(l)}r^{2}\dot{\phi}}{\dot{r}^{2}+r^{2}\dot{\phi}^{2}}.\label{eq:Ql-phi-gen-no-r}
\end{align}

\subsection{Specific Form of Generalized Forces}

As mentioned before, the generalized forces we consider in the model correspond to dynamical friction, $Q_{j}^{(\mathrm{DF})}$, and two leading PN corrections, namely, the conservative 1PN correction to the orbit and the leading gravitational wave emission term, which is the 2.5PN correction $Q_{j}^{(\mathrm{GW})}=Q_{j}^{(2.5\mathrm{PN})}$.  

\subsubsection{Post-Newtonian Corrections}

Following~\cite{Will:2011nz}, we write the PN corrections to the relative acceleration of the two bodies perturbatively as 
\begin{equation}
\frac{d\mathbf{v}}{dt}=  \frac{Gm}{r^{2}}\left\{ \frac{1}{c^{2}}\mathbf{A}_{1\mathrm{PN}}+\frac{1}{c^{4}}\mathbf{A}_{2\mathrm{PN}}+\frac{1}{c^{5}}\mathbf{A}_{2.5\mathrm{PN}}+\frac{1}{c^{6}}\mathbf{A}_{3\mathrm{PN}}+\frac{1}{c^{7}}\mathbf{A}_{3.5\mathrm{PN}}+\ldots\right\},
\end{equation}
with $c$ being the speed of light in vacuum.

The 1PN term contribution above is of the form~\cite{Will:2011nz}
\begin{equation}
\mathbf{A}_{1\mathrm{PN}}=\left\{ (4+2\eta)\frac{Gm}{r}-(1+3\eta)v^{2}+\frac{3}{2}\eta\dot{r}^{2}\right\} \hat{\mathbf{r}}+(4-2\eta)\dot{r}v\mathbf{\hat{v}},
\end{equation}
where $\eta = \mu/m$. Assuming the masses are time-independent, the corresponding 1PN force becomes
\begin{align}
\mathbf{F}^{(1\mathrm{PN})}= & \mu\frac{d\mathbf{v}}{dt}\bigg|_{1\mathrm{PN}}\nonumber \\
= & \frac{Gm\mu}{r^{2}c^{2}}\left[(4+2\eta)\frac{Gm}{r}-(1+3\eta)v^{2}+\frac{3}{2}\eta\dot{r}^{2}\right]\hat{\mathbf{r}} +\frac{Gm\mu}{r^{2}c^{2}}(4-2\eta)\dot{r}v\mathbf{\hat{v}}\label{eq:F1PN}.
\end{align}
To find its corresponding generalized force, we replace \eqref{eq:F1PN}
in (\ref{eq:QAr}) to obtain
\begin{equation}
Q_{r}^{(1\mathrm{PN})}= \frac{Gm\mu}{r^{2}c^{2}}\left[(4+2\eta)\frac{Gm}{r}-\left(3\eta+1\right)r^{2}\dot{\phi}^{2}+\dot{r}^{2}\left(-\frac{7}{2}\eta+3\right)\right].\label{eq:Qr-1PN}
\end{equation}
Likewise, from \eqref{eq:F1PN} and \eqref{eq:QAphi}, we obtain
\begin{equation}
Q_{\phi}^{(1\mathrm{PN})}=(4-2\eta)\frac{Gm\mu}{c^{2}}\dot{r}\dot{\phi}.\label{eq:Qphi-1PN}
\end{equation}
As stated before, the leading radiation term~\cite{blanchetPN} is given by the 2.5PN correction, $\mathbf{A}_{2.5\mathrm{PN}}$. Again, assuming the masses are time-independent, the gravitational radiation term takes the form~\cite{Will:2011nz} 
\begin{equation}
\mathbf{A}_{2.5\mathrm{PN}}=-\frac{8}{15}\eta\frac{Gm}{r}\left\{ \left(9v^{2}+17\frac{Gm}{r}\right)\dot{r}\hat{\mathbf{r}}-\left(3v^{3}+9\frac{Gm}{r}v\right)\hat{\mathbf{v}}\right\} ,
\end{equation}
which results in 
\begin{align}
\mathbf{F}^{(2.5\mathrm{PN})}= & \mu\frac{d\mathbf{v}}{dt}\bigg|_{2.5\mathrm{PN}}\nonumber \\
= & -\frac{8}{15}\eta\frac{G^{2}m\mu}{r^{3}c^{5}}\left[\left(9v^{2}+17\frac{Gm}{r}\right)\dot{r}\hat{\mathbf{r}}-\left(3v^{3}+9\frac{Gm}{r}v\right)\hat{\mathbf{v}}\right].
\end{align}
Using the above in (\ref{eq:QAr}) yields the radial part of the corresponding generalized force as
\begin{equation}
Q_{r}^{(2.5\mathrm{PN})}=-\frac{8}{15}\eta\frac{G^{2}m^2\mu}{r^{3}c^{5}}\dot{r}\left[6\left(\dot{r}^{2}+r^{2}\dot{\phi}^{2}\right)+8\frac{Gm}{r}\right].\label{eq:Qr-2.5PN}
\end{equation}
In the same way, using (\ref{eq:QAphi}), the angular part of this generalized force becomes
\begin{equation}
Q_{\phi}^{(2.5\mathrm{PN})}=\frac{8}{15}\eta\frac{G^{2}m^2\mu}{rc^{5}}\dot{\phi}\left(3\left(\dot{r}^{2}+r^{2}\dot{\phi}^{2}\right)+9\frac{Gm}{r}\right).\label{eq:Qphi-2.5PN}
\end{equation}

\subsubsection{Dissipation Due to Dynamical Friction}

The gravitational interaction between $m_{2}$ and its local dark
matter surrounding creates an effect known as dynamical friction (DF), similar in result to the electromagnetic
 friction, and slows down the rotating object, even though no electromagnetic or other forces in addition to gravity are assumed. This induces a dissipation
of energy from the system, whose instantaneous power loss can be modeled as~\cite{Eda:2014kra} 
\begin{equation}
P^{(\mathrm{DF})}=\frac{dE^{(\mathrm{DF})}}{dt}=-4\pi G^{2}\frac{m_{2}^{2}\rho_{\mathrm{DM}}(r)}{v}\xi(v)\ln\left(\Lambda\right).
\end{equation}
Here, $\rho_{\mathrm{DM}}(r)$ is the dark matter density profile, $v$ is the relative speed of the objects in polar
coordinates, and $\xi(v)=\gamma^2(1+ v^2/c^2)^2$ is a relativistic correction to the DF in which $\gamma=\sqrt{1- v^2/c^2}$ is the familiar special relativistic factor. The term $(1+ v^2/c^2)^2$ accounts for an increase in the deflection angle of the DM when considered as a collisional fluid due to the orbiting compact object, and $\gamma^2$ accounts for the relativistic momentum as seen by the compact object~\cite{Traykova:2021dua}. Furthermore, $\Lambda$ is called the Coulomb logarithm
and is defined as $\Lambda=b_{\mathrm{max}}v_{\mathrm{typ}}^{2}/G\mu$,
where $b_{\mathrm{max}}$ is the maximum impact parameter, and $v_{\mathrm{typ}}^{2}$
is the typical speed (squared) of the rotating object (or the relative speed squared in the CM frame). 
We take $\Lambda=3$ ~\cite{Eda:2014kra}.

Since this force is assumed to act only parallel to $\mathbf{v}$,
we can use the above formula in (\ref{eq:Ql-r-gen-no-r}) and (\ref{eq:Ql-phi-gen-no-r})
to obtain\footnote{Note that when $m_2\ll m_1$, $\mu \approx m_2$.}
\begin{align}
Q_{r}^{(\mathrm{DF})}= & \frac{P^{(\mathrm{DF})}\dot{r}}{\dot{r}^{2}+r^{2}\dot{\phi}^{2}}=-4\pi G^{2}m_{2}^{2}\xi(v)\rho_{\mathrm{DM}}(r)\frac{\dot{r}}{\left(\dot{r}^{2}+r^{2}\dot{\phi}^{2}\right)^{\frac{3}{2}}}\ln\left(\Lambda\right),\label{eq:Qr-DF}\\
Q_{\phi}^{(\mathrm{DF})}= & \frac{P^{(\mathrm{DF})}r^{2}\dot{\phi}}{\dot{r}^{2}+r^{2}\dot{\phi}^{2}}=-4\pi G^{2}m_{2}^{2}\xi(v)\rho_{\mathrm{DM}}(r)\frac{r^{2}\dot{\phi}}{\left(\dot{r}^{2}+r^{2}\dot{\phi}^{2}\right)^{\frac{3}{2}}}\ln\left(\Lambda\right).\label{eq:Qphi-DF}
\end{align}

\subsection{Lagrangian Equations of Motion}

We now have all the information to write down the Lagrangian equations
of motion (\ref{eq:EL-Gen-1}) and (\ref{eq:EL-Gen-2}), which will take the form (with Newtonian parts on the left and non-Newtonian ones on the right)
\begin{align}
\frac{d}{dt}\left(\frac{\partial L}{\partial\dot{r}}\right)-\frac{\partial L}{\partial r} = & Q_{r}^{(1\mathrm{PN})} + Q_{r}^{(2.5\mathrm{PN})}+Q_{r}^{(\mathrm{DF})},\\
\frac{d}{dt}\left(\frac{\partial L}{\partial\dot{\phi}}\right)-\frac{\partial L}{\partial\phi} = & Q_{\phi}^{(1\mathrm{PN})} + Q_{\phi}^{(2.5\mathrm{PN})}+Q_{\phi}^{(\mathrm{DF})}.
\end{align}
Replacing (\ref{eq:L}), (\ref{eq:V-Newton}), (\ref{eq:Qr-1PN}),
(\ref{eq:Qphi-1PN}), (\ref{eq:Qr-2.5PN}), (\ref{eq:Qphi-2.5PN}),
(\ref{eq:Qr-DF}), and (\ref{eq:Qphi-DF}) in the above yields
\begin{align}
\ddot{r}-r\dot{\phi}^{2}+\frac{Gm}{r^{2}}= & -\frac{Gm}{c^{2}r^{2}}\left[\frac{16G\mu\dot{r}}{5c^{3}r}+3\eta+1\right]\left(\dot{r}^{2}+r^{2}\dot{\phi}^{2}\right)\nonumber \\
 & -\frac{G\dot{r}}{c^{2}r^{2}}\left(\frac{64G^{2}\mu m^{2}}{15c^{3}r^{2}}-4m\dot{r}+\frac{\mu\dot{r}}{2}\right)+\frac{2G^{2}m}{c^{2}r^{3}}\left(2m+\mu\right)\nonumber \\
 & -\frac{4\pi G^{2}m_{2}^{2}\xi\rho_{\text{DM}}\ln(\Lambda)\dot{r}}{\mu\left(\dot{r}^{2}+r^{2}\dot{\phi}^{2}\right)^{3/2}},\label{eq:EoM-r-instant-2}
\end{align}
and
\begin{align}
r^{2}\ddot{\phi}+2r\dot{r}\dot{\phi}= & +\frac{8G^{2}\mu m\dot{\phi}}{5c^{5}r}\left(\dot{r}^{2}+r^{2}\dot{\phi}^{2}\right)+\frac{2G\dot{r}\dot{\phi}}{c^{2}}\left(2m-\mu\right)\nonumber \\
 & +\frac{24G^{3}\mu m^{2}\dot{\phi}}{5c^{5}r^{2}}-\frac{4\pi G^{2}m_{2}^{2}\xi\rho_{\text{DM}}\ln(\Lambda)r^{2}\dot{\phi}}{\mu\left(\dot{r}^{2}+r^{2}\dot{\phi}^{2}\right)^{3/2}}\label{eq:EoM-phi-instant-2}
\end{align}
These coupled differential equations can be solved together numerically to yield the orbits. 

\section{Orbits and Gravitational Waves\label{Sec:Orbits-GWs}}
\label{sec:GWpatterns}

\subsection{Orbital Evolution}
\begin{figure}[H] 
  \includegraphics[width=0.85\textwidth]{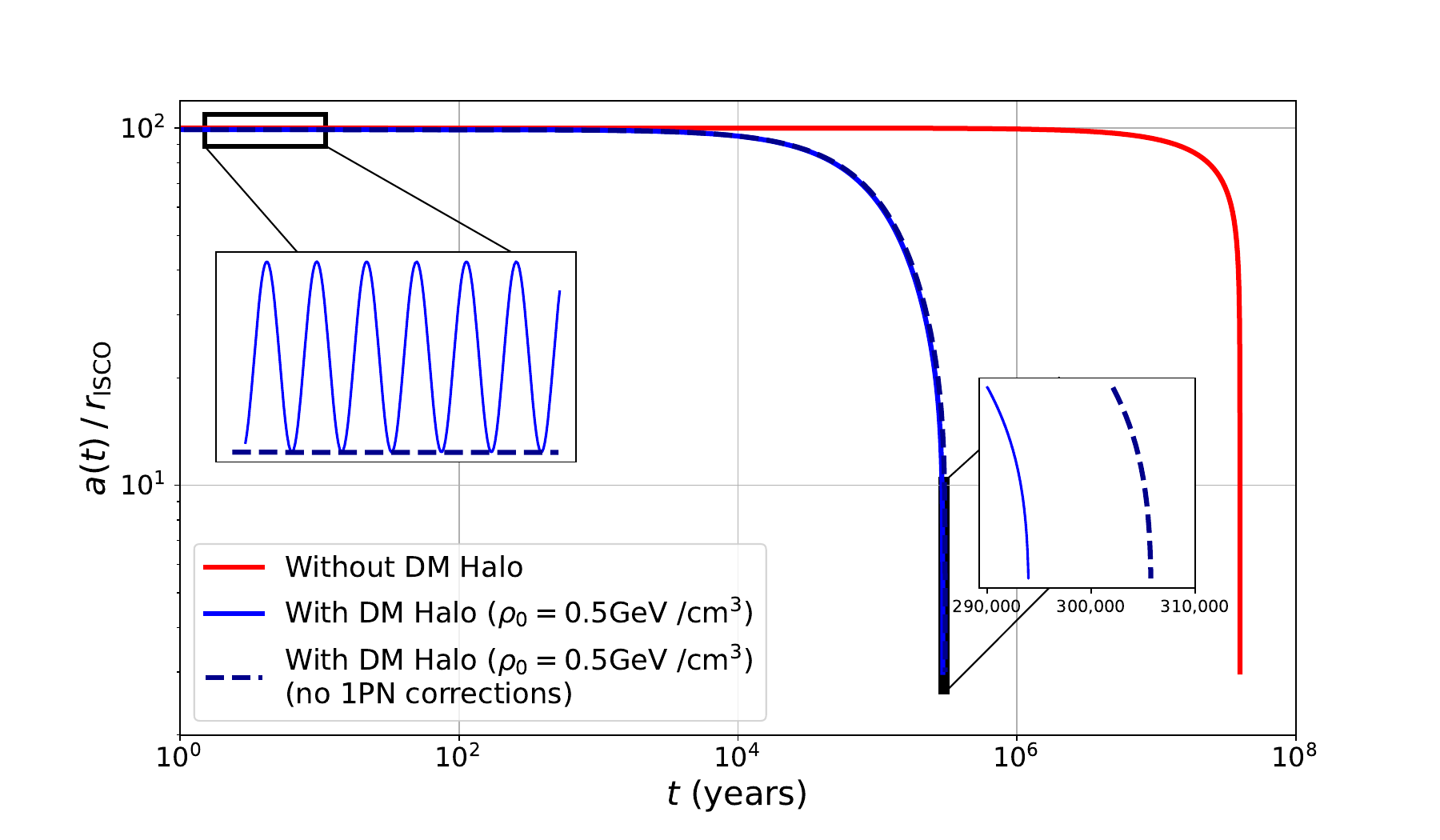} 
    \caption{Semi-major  axis of the orbit as a function of time, where in all cases, evolution ranges from $100 \;r_{\rm ISCO}$ to $3 \;r_{\rm ISCO}$. The dynamical friction term expedites energy dissipation and modifies the radial evolution of the binary. The no 1PN corrections curve refers to setting the $L_{1\text{PN}}$ terms to zero, which has a noticeable effect, especially at the latter stages of the inspiral. All the curves above include the GW dissipative term.}
  \label{fig:semimajor}
\end{figure}

For our given parameters, to illustrate the effect of these PN corrections and DM friction, the semi-major axis $a(t)$ was computed with the numerical solutions to (\ref{eq:EoM-r-instant-2}) and (\ref{eq:EoM-phi-instant-2}) as an evolution measure for all the effects introduced into the Lagrangian. These can be found in Figure \ref{fig:semimajor}. 

As expected, DM friction accelerates the inspiral by several orders of magnitude. Note, however, that the addition of the 1PN term into the Lagrangian introduces a different orbital evolution, and in the presence of DM, it accelerates the orbital decay due to the effects of dynamical friction.  This is particularly important during the later stages of the inspiral, where it differs from the regular GW radiative term. Note that the oscillations in the semi-major axis due to orbital precession from the 1PN term are still present at the later stages of the evolution; it is simply the scaling in plotting it alongside its no-1PN-corrections counterpart that makes the oscillations appear to vanish. Even though the 1PN term is a conservative term and would normally only introduce orbital precession, the presence of dark matter has affected the orbital evolution, which caused the orbit to lose energy faster than it would have with the normally used 2.5PN radiative term in the literature. One should expect that inclusion of higher order terms will also contribute to the inspiral's evolution. 

\subsection{Gravitational Wave Analysis}
\begin{figure}[H]
  \includegraphics[width=0.9\textwidth]{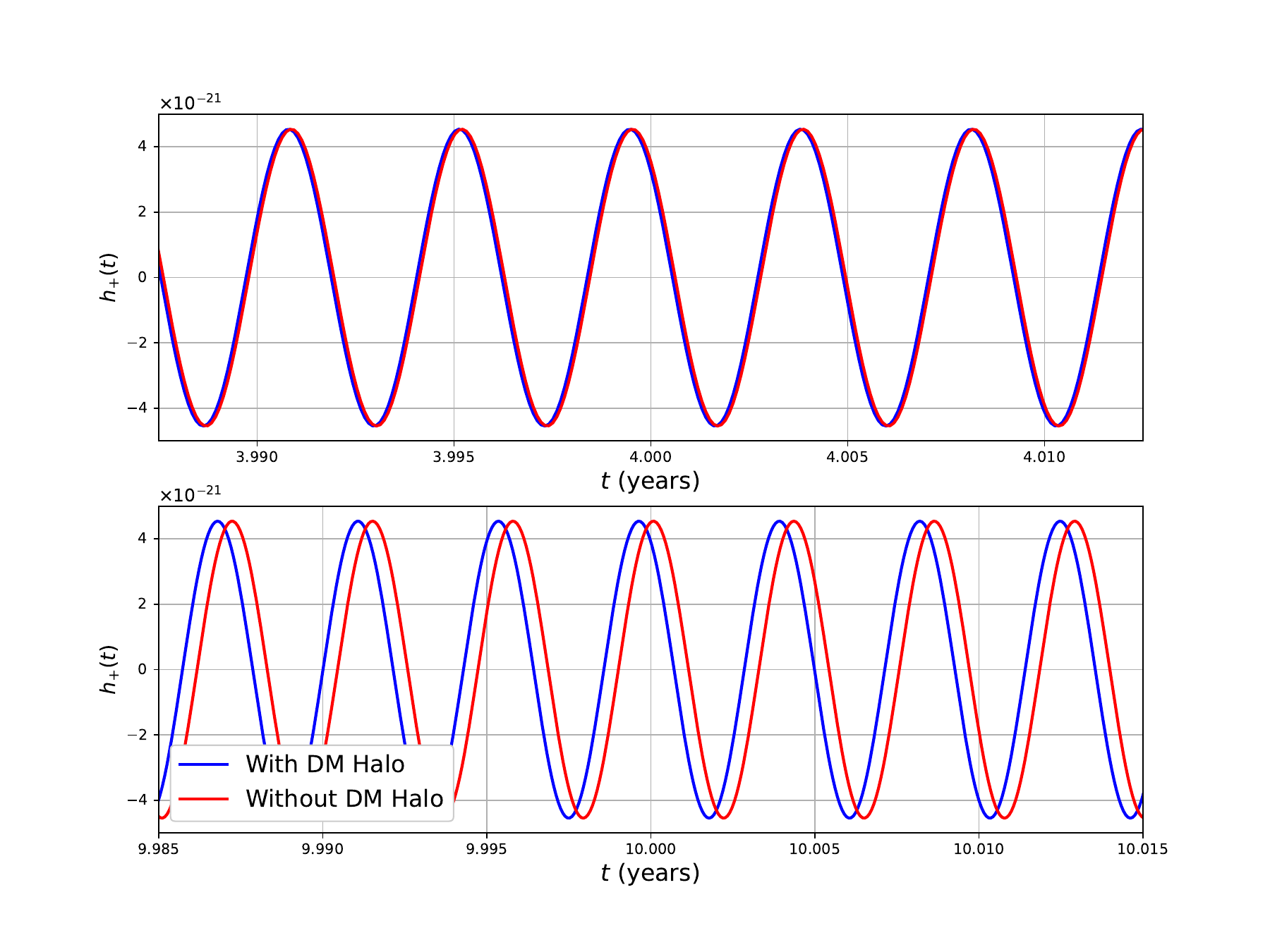} 
  \caption{Example  of the time-domain ``plus'' polarization waveform, $h_{+}(t)$, following the evolution of the $\rho_0 = 0.5$ GeV/cm$^3$ DM halo vs. vacuum with all PN corrections. Assuming the system enters in-band at the early stages of the inspiral ($70 \; r_{ \rm ISCO}$ at $t=0$) and at a point where the DM mini-spike is not at its densest, the timescale of this dephasing is set to start at about the four-year mark, which is within  LISA's lifetime. }
  \label{fig:waveform}
\end{figure}

Gravitational wave observatories measure the strain of the waves,
which is the Fourier mode $h_{\sigma}$ of the perturbations $h_{ij}$,
where $\sigma=+,\,\times$ are the polarization of the waves. Having found the orbits by solving the equations of motions
\eqref{eq:EoM-r-instant-2} and \eqref{eq:EoM-phi-instant-2}, we replace the resulting
$r(t)$ and $\phi(t)$ in
\begin{align}
x(t)= & r(t)\cos\left[\phi(t)\right],\\
y(t)= & r(t)\sin\left[\phi(t)\right],
\end{align}
and use the Cartesian Dirac deltas to express the matter density of
the system as
\begin{equation}
\rho(\mathbf{x},t)=\rho_{\mathrm{DM}}(r)+\mu\delta(x-x(t))\delta(y-y(t))\delta(z).
\end{equation}
Using these, we can find the quadrupole moment tensor
\begin{equation}
M^{ij}=\int x^{i}x^{j}\rho(t,\mathbf{x})\,d^{3}x,
\end{equation}
where $i,j=1,2,3$ correspond to the $x,\,y,\,z$ coordinates, respectively.
Finally, we can use $M^{ij}$ to compute the plus and cross polarizations
of the gravitational wave strains using 
\begin{align}
h_{+}(t;\bar{\theta},\bar{\phi})= & \frac{1}{R}\frac{G}{c^{4}}\left[\ddot{M}_{11}\left(\cos^{2}\left(\bar{\phi}\right)-\sin^{2}\left(\bar{\phi}\right)\cos^{2}\left(\bar{\theta}\right)\right)\right.\nonumber \\
 & +\ddot{M}_{22}\left(\sin^{2}\left(\bar{\phi}\right)-\cos^{2}\left(\bar{\phi}\right)\cos^{2}\left(\bar{\theta}\right)\right)\nonumber \\
 & -\ddot{M}_{33}\sin^{2}\left(\bar{\theta}\right)-\ddot{M}_{12}\sin\left(2\bar{\phi}\right)\left(1+\cos^{2}\left(\bar{\theta}\right)\right)\nonumber \\
 & \left.+\ddot{M}_{13}\sin\left(\bar{\phi}\right)\sin\left(2\bar{\theta}\right)+\ddot{M}_{23}\cos\left(\bar{\phi}\right)\sin\left(2\bar{\theta}\right)\right]
\end{align}
and 
\begin{align}
h_{\times}(t;\bar{\theta},\bar{\phi})= & \frac{1}{R}\frac{G}{c^{4}}\left[\left(\ddot{M}_{11}-\ddot{M}_{22}\right)\sin\left(2\bar{\phi}\right)\cos\left(\bar{\theta}\right)\right.\nonumber \\
 & +2\ddot{M}_{12}\cos\left(2\bar{\phi}\right)\cos\left(\bar{\theta}\right)-2\ddot{M}_{13}\cos\left(2\bar{\phi}\right)\sin\left(\bar{\theta}\right)\nonumber \\
 & \left.+2\ddot{M}_{23}\sin\left(2\bar{\phi}\right)\sin\left(\bar{\theta}\right)\right],
\end{align}
where $R$ is the distance of the observatory to the CM of the binary system, and $\bar{\theta},\,\bar{\phi}$ are related to the
relative orientation of the frames of reference of the source and
the observer~\cite{maggiore2018}. Note that these are different from $\theta$ and $\phi(t)$, which denote the angular position coordinates of the binary in the CM frame. A benefit to this approach is that by directly solving the Lagrangian equations of motion, one can freely compute the explicit waveforms as a function of time, where we can observe the rapid effect from DM friction through the dephasing of the waveforms. For our purposes, we have considered a source at \mbox{$ \{ R, \bar{\theta}, \bar{\phi} \}  = \{ 1 \text{Mpc}, \; 0,\; 0 \} $}. Note that with this choice of angles, $h(t) = h_{+}(t)$. The result is plotted in Figure \ref{fig:waveform}.

\begin{figure}[H]
  \includegraphics[width=0.9\textwidth]{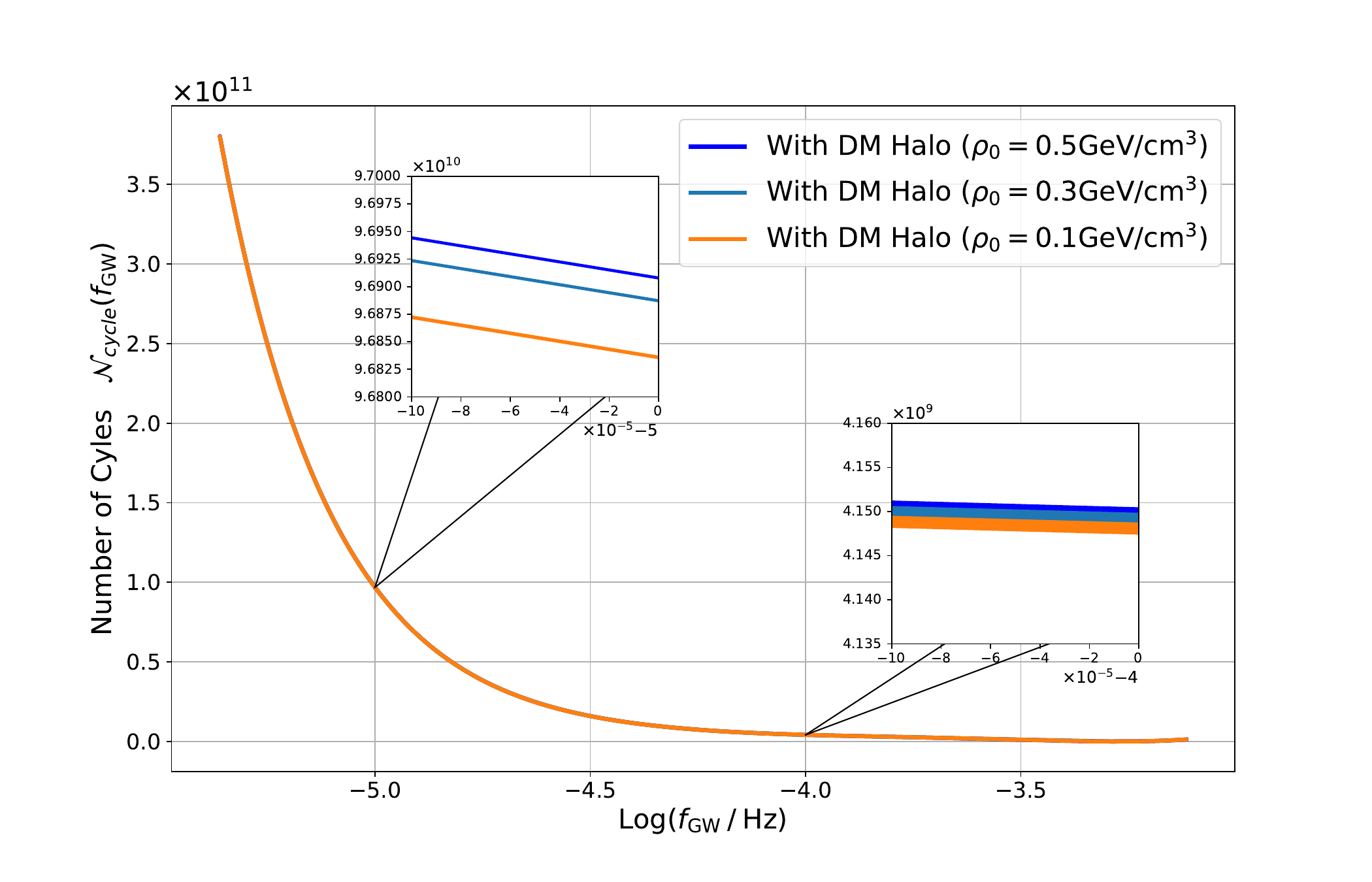} 
  \caption{Number  of cycles comparison for different DM densities and vacuum as a function of GW frequency. Inset plots represent zoomed-in sections of the main plot at $10^{-5}$ Hz (left inset) and $10^{-4}$~Hz (right inset).}
  \label{fig:Ncycle}
\end{figure}

To fully encapsulate this GW dephasing effect, we analyze the number of cycles, $\mathcal{N}_{\text {cycle }}$, that the system can be in-band as a function of the gravitational wave frequency, $f_{GW}$. Note that for this, we may use the relations that come directly from solving the Lagrangian equations of motion of the phase $\phi(t)$, which is related to the GW phase with $\phi_{GW}(t) = 2 \phi(t)$, and the number of cycles would simply be
\begin{align}
    \mathcal{N}_{\text{cycle }}(t) \equiv \frac{\phi_{\text{No DM}}(t) - \phi_{\text{DM}}(t)}{\pi}.
\end{align}
Accompanied by the following relation,
\begin{align}
    \dot{\phi}(t) = 2 \pi f_{GW}(t) \,,
\end{align}
one may obtain $\mathcal{N}_{\text {cycle }}(f_{GW})$ with the help of the solutions to (\ref{eq:EoM-r-instant-2}) and (\ref{eq:EoM-phi-instant-2}). We can see the result in Figure \ref{fig:Ncycle}.

\vspace{-3pt}

\begin{figure}[H]
  \includegraphics[width=0.85\textwidth]{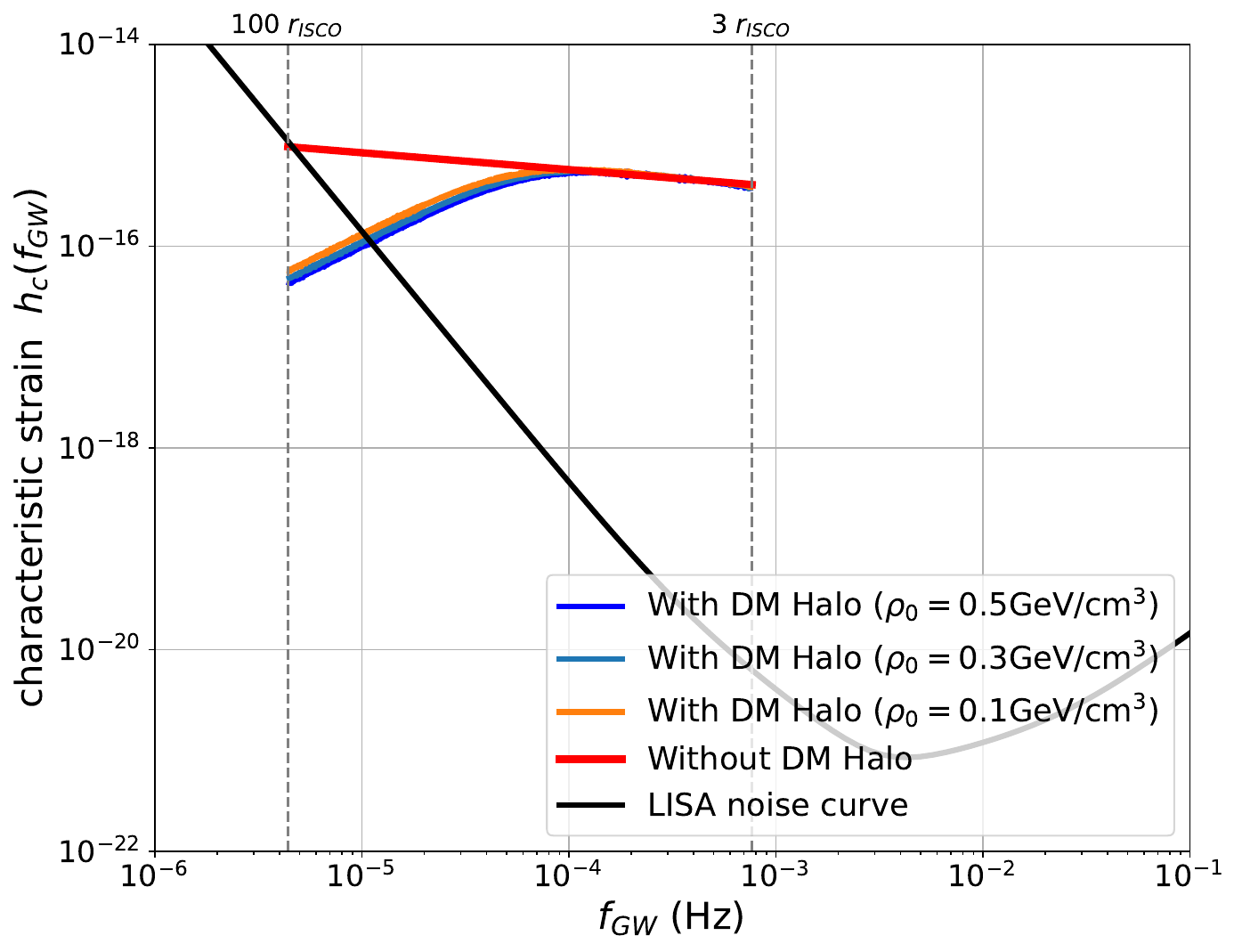} 
  \caption{The strain and detectability for LISA's sensitivity curve given for various DM densities parametrized by $\rho_0$. GW energy dissipation is dominant over DM friction during the later stages of the inspiral, which can be seen by the agreement of all strains at higher frequencies. Thus, searches for this DM friction would be the most sensitive during earlier parts of the inspiral but not earlier than the strain curves crossing LISA's sensitivity. For our choice of parameters, this occurs near $10^{-5}$~Hz. } 
  \label{fig:freqstrain}
\end{figure}

The number of cycles in the presence of a DM halo will be lower than in a vacuum, which is expected as the DM friction would translate a faster inspiral into fewer orbits, being in-band for GW emission. What is important for the detection of a DM spike from DM friction would be the particular shape of the waveform in frequency space, as interferometers set out to find signals from match-filtering techniques in the spectral domain. Hence, it is more illustrative to work in the frequency domain. In particular, we are interested in the detectability of these results with the LISA sensitivity curve, as well as the different signatures obtained from changing DM parameters. To adequately compare this to the detector's sensitivity to estimate signal-to-noise ratios, we compute the characteristic strain $h_c(f_{GW})$, given by
\begin{align}
    h_c(f_{GW}) = 2 f_{GW} | \tilde{h} (f_{GW}) | 
\end{align}
with $\tilde{h} (f_{GW})$ being the Fourier transform of the GW time signal. The numerical computation of the frequency domain in the case of the system we chose compared to the LISA sensitivity curve can be seen in Figure \ref{fig:freqstrain} (see also Appendix \ref{appx}). We can see there is a clear distinction between the cases with DM halo and the case without DM halo in lower frequencies, which will be observable by LISA. 

\begin{figure}[H]
  \includegraphics[width=0.85\textwidth]{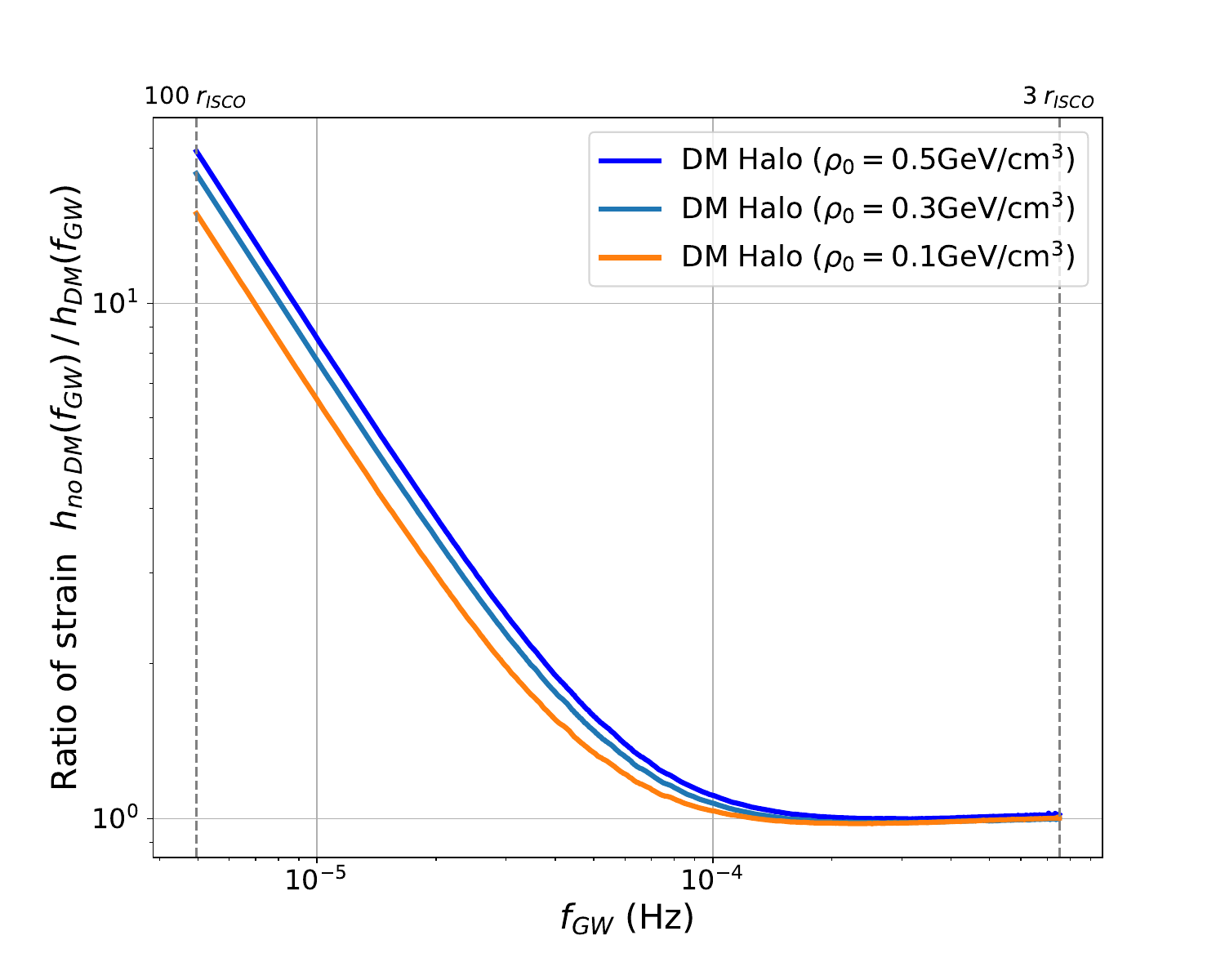} 
  \caption{Ratio between the characteristic strains between the case without dark matter, and the different values of $\rho_{0}$. As expected, the largest deviations occur during the early stages of the inspiral. Eventually, dissipation through gravitational wave radiation dominates dark matter friction and leads all waveforms to converge on the same frequency evolution. This stage, although the brightest on the LISA band, has the smallest contributions from dark matter signatures.  } 
  \label{fig:ratio}
\end{figure}

To further accentuate these effects, looking at the ratio between the characteristic strain without any DM, or $h_{no \: DM}(f_{GW})$, and the characteristic strain with dark matter, $h_{ DM}(f_{GW})$, we can see the large deviations mostly occur at the early stages of the inspiral, as shown in Figure \ref{fig:ratio}. In particular, we are referring to the inspiral stages between $\sim 4 \cdot 10^{-6}$~Hz to $10^{-4}$~Hz, or $100r_{ISCO}$ to $\sim 12 r_{ISCO}$. This should be expected, as energy dissipation through GWs increases as the orbit shrinks. Eventually, GW emission dominates the energy loss from dynamical friction, and all profiles converge to the strain without dark matter. This, of course, means that the range of detectability of these DM environmental effects solely depend on the stage of the binary, where the signal would be maximal at lower frequencies, where clearly they will differ by at least an order of magnitude, as seen in Figure \ref{fig:ratio}.

\vspace{-6pt}

\section{Discussion and Conclusions\label{sec:conclusion}}
\label{sec:discussion}

Compact binary systems with an environment around them, particularly a DM spike, are a valuable experimental arena for the potential detection of effects associated with new physics, particularly the detection of the effects of DM on the emitted gravitational waves from such systems. Therefore, it is important to have a robust theoretical framework that can incorporate as many physical effects as possible, particularly general relativistic and DM effects. In this paper, we have employed a Lagrangian method that allows for the energetically consistent incorporation of conservative and dissipative effects, in particular, post-Newtonian corrections to the orbit up to any desired known order, as well as a relativistic dynamical friction term with DM. This allows us to expand the range of validity of the orbital evolution throughout the later stages of the inspiral. In particular, we applied this method to the case of a binary with a dark matter spike, where in addition to a relativistic form of dynamical friction, we have considered the leading PN corrections to the gravitational waveform, i.e., 1PN and 2.5PN terms, corresponding to the orbital precession and gravitational wave dissipation, respectively. After obtaining the analytical Lagrangian equations of motion, they were solved numerically to yield the orbits. These solutions were then used to compare the orbits in cases with and without DM halo and PN corrections. We have shown that the conservative 1PN term, while by itself will not lead to any orbital energy loss, in the presence of a DM spike, it will modify the evolution of the orbit, where in a wealth of literature, the dominant terms included are the GW dissipation term (equivalent to the 2.5PN term). Hence, any future realistic analysis should require corrections in order to improve the detectability of these effects. Furthermore, using the orbit equations, the modified GW waveform emitted from the binary in the presence of a DM spike was derived, which shows a dephasing effect compared to the no-DM case.  Moreover, we have computed the frequency space strain of the GWs for three different DM densities and compared them against the LISA frequency sensitivity curve. Our calculations show that the modification effects resulting from the presence of a DM spike are observable by LISA. 

In this method, the addition of further physical effects can be carried out in an energetically consistent manner through the Lagrangian and generalized forces. We have included a relativistic dynamical friction term and spike profile, as well as shown how to account for the addition of post-Newtonian corrections up to any known order and, in particular, corrections to both the orbital motion and GW emission up to the recently derived higher order terms~\cite{https://arxiv.org/pdf/2304.13647.pdf}. Therefore, utilizing this framework could be quite beneficial in the search for DM effects, since following a potential inspiral beyond the Newtonian regime of the evolution will provide better GW templates for match filtering methods with the LISA detector or other observatories such as Cosmic Explorer. Moreover, this provides an alternative modeling tool to compare the current or future work in GWs sourced by DM~\cite{Speeney:2022ryg, bertoneDarkMatterMounds2024}. One may also imagine modifying the Lagrangian to add other effects, such as the accretion of DM or baryonic matter by the binary, which can be incorporated into our framework quite easily by adding a time dependence on the masses. Lastly, this framework also provides the liberty to modify the environment density $\rho(r)$ to include dynamical and companion interaction effects $\rho(r) \to \rho(t, r, \phi)$.


\begin{acknowledgments}
We thank Gianfranco Bertone for valuable discussions and input. This work has emerged as a follow-up from the international innovative teaching and research program EXPLORE (``EXPeriential Learning Opportunity through Research and Exchange'') of York University, the University of Alberta, and Goethe University Frankfurt. We thank all mentors, junior mentors and participants of EXPLORE for stimulating discussions and a fruitful collaboration. The EXPLORE program is supported by the Academic Innovation Fund (AIF) at York University, Goethe University QSL (``Quality Assurance in Teaching'') funds, the State of Hesse within the Research Cluster ELEMENTS (Project ID 500/10.006), and the Deutsche Forschungsgemeinschaft (DFG, German Research Foundation) through the CRC-TR 211 ``Strong-interaction matter under extreme conditions''-- project number 315477589-TRR 211.
S. R. acknowledges the support of the Natural Science and Engineering Research Council of Canada, funding reference No. RGPIN-2021-03644 and No. DGECR-2021-00302. 
\end{acknowledgments}


\appendix
\section{Analytical Approximation of the Characteristic Strain}\label{appx}  

We may attempt to compute the strain $h_c(f)$ directly using the stationary phase approximation (SPA):
\begin{align}
    \tilde{h}_{+, \times} (f) = \int h_{+, \times} (t) \; e^{ 2  \pi i ft} \; dt \:  \approx \: \frac{1}{2} A_{+,\times} (f) \; e^{ i \Psi_{+,\times}(f) }, 
\end{align}
where
\begin{align}
    A_{+}(f) = A_0(f) \; \Big( \frac{1 + \cos^2{(\bar{\theta}})} {2} \Big),\quad   A_{\times}(f) = A_0(f) \; \Big( \cos{(\bar{\theta})} \Big),\label{Af}
\end{align}
and
\begin{equation}
A_{0}(f) = \frac{ 4 \; (G M)^{5/3} \; (\pi f)^{2/3}}{ R \; c^4} \Big( \frac{ 2 \pi} { \ddot{\phi} (f)} \Big)^{1/2},
\end{equation}
and
\begin{align}
    \Psi_{+}(f) = 2 \pi f t_c - \phi_c - \frac{\pi}{4} + \phi(f), \quad \Psi_{\times}(f) = \Psi_{+}(f) + \frac{\pi}{2}.\label{PSIf}
\end{align}
Here, $t_c$ and $\phi_c$ are time and phase at coalescence, respectively. We should note that solving (\ref{eq:EoM-r-instant-2}) and (\ref{eq:EoM-phi-instant-2}) in time provide us with $\phi(t)$ to insert in both (\ref{Af}) and (\ref{PSIf}). This can be related to the frequency domain with the time, frequency, and orbital phase relation
\begin{equation}
    \dot{\phi}(t) = 2 \pi f(t).
\end{equation}
This approach, however, is not immediately useful as $\ddot{\phi}(f)$ will be an oscillatory function of $f$, which can take negative values due to the introduction of precession from the $1$PN term. Hence, the complete Fourier transform has to be computed numerically.

\bibliographystyle{apsrev4-2}

\bibliography{GWDMmain}

\end{document}